\documentclass[%
 reprint,
 amsmath,amssymb,
 aps,
pra,
]{revtex4-2}

\usepackage{amsmath}
\usepackage{verbatim}
\usepackage{graphicx}
\usepackage{physics}
\usepackage{dcolumn}
\usepackage{hyperref}
\usepackage{subfigure}
\usepackage{float}
\usepackage{bm}

\begin{document}

\preprint{APS/123-QED}

\title{Detection of weak magnetic fields using nitrogen-vacancy centers with maximum confidence}

\author{Ghazi Khan}
\affiliation{School of Science \& Engineering, Lahore University of Management Sciences (LUMS), Opposite Sector U, D.H.A, Lahore 54792, Pakistan}

\author{Adam Zaman Chaudhry}
\email{adam.zaman@lums.edu.pk}
\affiliation{School of Science \& Engineering, Lahore University of Management Sciences (LUMS), Opposite Sector U, D.H.A, Lahore 54792, Pakistan}


\date{\today}

\begin{abstract}
The problem of detection of magnetic fields using NV centers, that is, to check whether a weak magnetic field is present or not, can be tackled using quantum state discrimination theory. In this regard, we find the POVMs that maximize the confidence of any given measurement, taking the interaction time as well as decoherence into account. We apply our formalism over a wide range of scenarios encompassing constant and oscillating magnetic fields, while also using techniques such as dynamical decoupling to improve the confidence and extending our treatment to ensembles of NV centers. 
\end{abstract}

\maketitle


\section{\label{sec:level1}Introduction\protect}
Quantum sensing provides us the opportunity to exploit quantum coherence towards detecting weak signals with an accuracy and precision that approaches the fundamental limits \cite{budker2007optical}. In this regard, the detection of weak magnetic fields is an important problem in diverse fields such as data storage, biomedical sciences, material science and quantum control \cite{freeman2001advances,grinolds2011quantum,greenberg1998application}. Towards this end, the use of NV center as magnetic field sensors was first proposed by Refs.~\cite{degen2008scanning,taylor2008high}. This was then demonstrated using single NV centers \cite{maze2008nanoscale,balasubramanian2008nanoscale} and eventually for ensembles \cite{acosta2009diamonds}. Consequently, the small size, impressive magnetic field sensitivity, robustness in diverse environments, and wide-range-temperature operation of NV centers has made it a key player in developing the temporal profiles of weak magnetic fields \cite{pham2012enhanced,horowitz2012electron,kagami2011detection,de2011single,mcguinness2011quantum,laraoui2010magnetometry,hall2009sensing,balasubramanian2009ultralong,chang2008mass}. 

The typical approach to NV center magnetometry is to initialize a superposition state of two different energy levels. This superposition state then develops a phase difference in the presence of a magnetic field. The phase difference depends on the parameters of the magnetic field, which can thus be read out to determine the unknown signal. However, our paper takes a different approach to this problem. We do not wish to measure the specific parameters of the magnetic field, rather we wish to determine whether or not the field, about which we have some prior information, is present or not. Towards this goal, our basic idea again involves preparing the NV center in an equal superposition state of two energy levels $\ket{0}$ and $\ket{1}$, leading to the quantum state $\rho_0$. The Hamiltonian describing the interaction of the NV center with the magnetic field $B(t)$ is $H(t)=\pi\gamma B(t) \sigma_{z}$ where $\sigma_{z}$ is the standard Pauli matrix and $\gamma=28$ Hz/nT. In the absence of decoherence, the state remains the same if the magnetic field is absent. However, in the presence of the magnetic field, the state becomes changes to $\rho_1$ which depends on the magnetic field. Consequently, the problem reduces to simply discriminating between the states $\rho_0$ and $\rho_1$ in order to detect the magnetic field.

The theory of quantum discrimination has received considerable attention in the field of quantum cryptography, with many approaches having being developed such as minimum-error discrimination (ME) \cite{helstrom1976quantum,herzog2004minimum,holevo1982probabilistic,yuen1975optimum,ha2013complete,ha2014discriminating}, unambiguous discrimination (UD) \cite{peres1988differentiate,ivanovic1987differentiate,dieks1988overlap,jaeger1995optimal}, maximal confidence (MC) \cite{croke2006maximum}, error margin (EM) methods  \cite{sugimoto2009discrimination,hayashi2008state,touzel2007optimal}, and fixed rate of inconclusive results (FRIR) methods \cite{ha2017optimal,fiuravsek2003optimal,zhang1999general,chefles1998quantum}. Here our objective is to use quantum discrimination to distinguish between two states in order to infer the presence of a magnetic field. Our proposed scheme aims to maximize the confidence of our measurement by optimizing the time our NV centers are allowed to interact with the magnetic field. This formalism is applied both constant and fluctuating magnetic field profiles. We then extend our treatment to consider ensembles of NV centers. Finally we take a look at how such generalized measurements can actually be implemented.

This paper is organised as follows. In \autoref{sec:2} we discuss physical properties of the nitrogen-vacancy that are going to be used as detectors.In \autoref{sec:3} the formalism for quantum state discrimination is discussed alongside the solutions to finding optimal generalised operators. In Section IV we use our formalism towards the detection of static and oscillating magnetic fields in single NV centers followed by NV center ensembles. Finally, in Section V, the physical implementation of these generalised operators is discussed. We conclude in Section VI.

\section{\label{sec:2}NV Center\protect}
As we have highlighted in the introduction, detection of a magnetic field using a NV center is possible due to the phase difference picked up by the quantum state of the NV center in the presence of a magnetic field. Unfortunately, several mechanisms contribute to the dephasing of the NV spin. These dephasing mechanisms are dominated by the following:
\begin{itemize}
  \item Paramagnetic substitutional nitrogen impurities (P1 centers, \(S=1/2\)), resulting from the NV$^-$ enriching process, effectively create an electronic spin bath that couples to the NV spins via incoherent dipolar interactions. Such decoherence is typically dominant for type-Ib samples.
  \item $^{13}$C nuclei (\(I=1/2\)) also form a typical nuclear spin bath which is a source for considerable NV spin dephasing. This is especially in type-IIa samples which have much smaller nitrogen concentrations.
  \item Miscellaneous sources of non-magnetic noise sources such as temperature fluctuations, electric field noise, magnetic field gradients, and inhomogeneous strain can also significantly impact dephasing times depending on the sample and the experimental setup.
\end{itemize}
A common approach to model the effect of the environment on the NV center is via a classical Gaussian noise field \(B_d(t)\) \cite{wang2012comparison}. This field is modeled by an Ornstein-Uhlenbeck stochastic process with zero mean and a correlation function \(\langle B_d(0)B_d(t)\rangle=\kappa^{2} e^{-|t|/\tau_{c}}\), where $\tau_{c}$ is the correlation time of the bath and $\kappa$ is the coupling strength of the bath to the spin. For the case of a single NV center undergoing free decay we have $\langle \exp(-i\int_{0}^{T}B_{d}(t)dt)\rangle$ where $\langle\cdots\rangle$ indicates an average over different noise realisations; this gives us an exponential dephasing factor of $e^{-\kappa_{\text{single}}^{2}T^{2}/2}$. Equivalently, the decoherence factor is characterized by 
\begin{eqnarray}
\frac{1}{e^{-(t/T_{2,\text{single}}^{*})^{p}}},
\label{T2deco}
\end{eqnarray}
where
\begin{eqnarray}
T_{2,single}^{*}=\left[\frac{2}{\kappa_{\text{single}}^{2}}\right]^{1/2}
\textrm{and}
\;\;
p=2.
\end{eqnarray}
In order to evaluate the decoherence for ensembles of NV centers we need to integrate over a distribution of various $\kappa$ and $\tau_{c}$. With a full derivation available in Ref.~\cite{bauch2020decoherence}, the result for the case of dipolar couplings to a spin bath results in the same expression as $T_{2,single}^{*}$ albeit with the stretched exponential parameter $p=1$ and $T_{2,ensemble}^{*}=1/\kappa_{ens}$. 

This approach of modeling the environment also maintains the flexibility to model NV centers in other environments. This means that while we limit ourselves to the choice of a NV center experiencing dipolar coupling to a surrounding electronic spin bath, which results in the stretched exponential parameter $p=1(2)$ for the ensemble (single) scenario, the parameter can be modified to model different environments. For example a non-integer value of $p$ might suggest the presence of strain gradients, magnetic fields, temperature gradients or other more complex dephasing mechanisms \cite{bauch2018ultralong}, these exact values of $p$ are well characterised for the single spin under a variety of environments \cite{de2009electron,maze2012free,hanson2008coherent}
Such modeling approaches have shown tremendous agreement for results involving theory and experiment in single NV centers \cite{maze2012free,de2009electron,dobrovitski2008decoherence,hall2014analytic} and ensembles \cite{bauch2018ultralong,bauch2020decoherence}. In short, by varying the noise parameters, the formalism we present for magnetic field detection can be applied to very different sets of experimental environments.

\section{\label{sec:3}Discrimination formalism for Magnetometry\protect}
Once a NV center has been prepared in a superposition state, it is clear that in the absence of a magnetic field, there is only dephasing with some time-dependent factor $\nu$. On the other hand, the presence of a magnetic field leads to evolution of the NV center with a phase factor $\mu$. We can already see then that our task for detection becomes quite simple where we have to simply discriminate between the two mixed states
\begin{eqnarray}
\begin{aligned}
\rho_{0}=\frac{1}{2}
 \left( \begin{array}{cc}
1 & \nu \\
\nu & 1
\end{array} \right),
\;\;
\rho_{1}=\frac{1}{2}
\left( \begin{array}{cc}
1 & \nu\mu \\
\nu\mu^{*} & 1
\end{array} \right).
\end{aligned}
\label{rhos}
\end{eqnarray}

The use of minimum error (ME) discrimination for detecting magnetic fields has already been explored in Ref.~\cite{chaudhry2015detecting}. However, such a discrimination technique is limited to only von-Neumann measurements. We follow an alternative strategy. We wish to maximize the confidence, that is, we maximize the chance that our state was in state $\rho_{0}(\rho_{1})$ given that the corresponding detector $\Pi_{0}(\Pi_{1})$ goes off. This, in general, requires generalized measurements.

Maximum confidence has been considered extensively in theory and experiment for solving various quantum information problems\cite{croke2006maximum,herzog2012optimized,mosley2006experimental,steudle2011experimental}. The strategy aims to discriminate between quantum states with maximum confidence in each conclusive
result, thereby keeping the probability of
inconclusive results as small as possible. The discrimination strategy proceeds as follows. We assume that we have been provided two mixed states $\rho_{0}$ and $\rho_{1}$, namely the states given in Eq.~\eqref{rhos}, which occur with some prior probability $\eta_{0}$ and $\eta_{1}$ respectively (here $\eta_{0}+\eta_{1}=1$). A complete measurement strategy is then described by three POVM operators $\Pi_{0}$, $\Pi_{1}$ and $\Pi_{?}$, which sum up to the identity operator $I_{d}$ in the $d$-dimensional joint Hilbert space $\mathcal{H}_{d}$. If $\Pi_{0}$ `clicks' we say that the state is $\rho_0$, if $\Pi_{1}$ `clicks' we say we have $\rho_{1}$, and if $\Pi_{?}$ clicks we say we have an inconclusive result. Our objective then is find the maximum confidence $C_{j}$. This is the conditional probability $P(\rho_{j}|j)$ that the state $\rho_j$ was indeed prepared give that the detector $\Pi_j$ clicked, that is,
\begin{eqnarray}
C_{j}=\frac{\eta_{j}\textrm{Tr}(\rho_{j}\Pi_{j})}{\textrm{Tr}(\rho\Pi_{j})},
\end{eqnarray}
with
\begin{eqnarray}
\rho=\eta_{0}\rho_{0}+\eta_{1}\rho_{1}.
\end{eqnarray}
We minimize the rate of inconclusive outcomes $P_{inc}$
\begin{eqnarray}
P_{inc}=\textrm{Tr}(\rho\Pi_{?}),
\label{inceq}
\end{eqnarray}
and we also have the usual POVM constraints
\begin{eqnarray}
\begin{aligned}
\Pi_{?}=I_{d}-\Pi_{1}-\Pi_{2}\geq0,\;\;\Pi_{1}\geq0,\;\;\Pi_{2}\geq0.
\end{aligned}
\label{summa}
\end{eqnarray}
For the case where we consider the two mixed states in Eq.~\eqref{rhos}, there is a general solution \cite{herzog2009discrimination}. We first find
\begin{eqnarray}
\widetilde{\rho_{0}}&&=\eta_1\rho^{-1/2}\rho_0\rho^{-1/2}\\\nonumber
&&=\gamma_{\text{max}}^{(0)}|\gamma_{0}\rangle \langle\gamma_{0}|+\gamma_{\text{min}}^{(0)}|\gamma_{1}\rangle\langle\gamma_{1}|,
\label{ann2}
\end{eqnarray}
where $\gamma_{max(min)}^{(0)}$ is the maximum (minimum) eigenvalue for $\widetilde{\rho_{0}}$ with the eigenvector $\ket{\gamma_0}(\ket{\gamma_1})$. The maximum confidence is then simply determined to be $C^{\text{max}}_0=\gamma^{0}_{\text{max}}$ and $C^{\text{max}}_2=1-\gamma^{(0)}_{\text{min}}$. The corresponding optimal probability for inconclusive results $P^{opt}_{inc}$ is found to be, while using the notation .
\begin{eqnarray}
P^{\text{opt}}_{\text{inc}} = \begin{cases} 
          1-\frac{\text{det}(\rho)}{\rho_{11}} & \text{if}\;\;\; |\rho_{01}|\geq\rho_{11}, \\
          1-\frac{\text{det}(\rho)}{\rho_{00}} & \text{if}\;\;\; |\rho_{01}|\geq\rho_{00},\\
          2|\rho_{01}| & \text{otherwise},
       \end{cases}
\end{eqnarray}
where $\rho_{ij}=\bra{\gamma_i}\rho\ket{\gamma_j}$.

Finally, the POVMs can be easily determined as rank one operators of the form $\Pi_{0}^{\text{opt}}=a_{\text{opt}}\ketbra{v}{v}$ and $\Pi_{1}^{\text{opt}}=b_{\text{opt}}\ketbra{w}{w}$, where
\begin{eqnarray}
          a_{\text{opt}}=1,\;\;\; & b_{\text{opt}}=0 &\;\; \text{if}\;\;\; |\rho_{01}|\geq\rho_{00}, \nonumber\\
          a_{\text{opt}}=0,\;\;\; & b_{\text{opt}}=1 &\;\; \text{if}\;\;\; |\rho_{01}|\geq\rho_{11},\nonumber\\
           a_{\text{opt}}=a_{0},\;\;\; & b_{\text{opt}}=b_{0}\;\; & \;\;\textrm{otherwise}.\nonumber
           \label{thirdcondition}
\end{eqnarray}
Here
\begin{eqnarray}
a_{0}=\frac{1-\frac{|\rho_{01}|}{\rho_{00}}}{1-\frac{|\rho_{01}|^{2}}{\rho_{00}\rho_{11}}},\;\;\;\;
b_{0}=\frac{1-\frac{|\rho_{01}|}{\rho_{11}}}{1-\frac{|\rho_{01}|^{2}}{\rho_{00}\rho_{11}}},
\label{abeq}
\end{eqnarray}
and $|v\rangle$ and $|w\rangle$ are the normalized states
\begin{eqnarray}
|v\rangle=\frac{\rho^{-1/2}|\gamma_{0}\rangle}{(\langle \gamma_{0}|\rho^{-1}|\gamma_{0}\rangle)^{1/2}},\;\;
|w\rangle=\frac{\rho^{-1/2}|\gamma_{1}\rangle}{(\langle \gamma_{1}|\rho^{-1}|\gamma_{1}\rangle)^{1/2}},
\label{an}
\end{eqnarray}
It must be noted that for the case where the third condition of Eq.~\eqref{thirdcondition} does not hold we then either have $a$ or $b$ end up as zero resulting in us having only two non-zero measurement projective von-Neumann measurement operators. We further note that as proven in Ref.~\cite{herzog2009discrimination}, this measurement would then simply correspond to the ME measurement.

\section{Detection of magnetic fields}
With our formalism presented, we can now proceed towards applying it towards magnetic field detection. Our objective is to find the operators $\Pi_1,\Pi_2 \textrm{ and } \Pi_?$ for some time $T$ that maximize the confidence $C_{j}^{max}$ while also keeping the $P_{inc}$ to a reasonable level. The intuition behind this makes sense, since for a small values of $T$ the phase accumulation is too small to discriminate between the two density matrices well, while for high values of $T$ the effect of decoherence is too strong.

\subsection{Single NV centers}
We first apply our formalism using a single NV center (\ref{T2deco}) with p=2. We use $\kappa= 3.6\, \mu \text{s}^{-1}$ and $\tau_{c}=25 \mu \text{s}$. These correspond to the free induction decay time of  $T_{2,single}^{*}=0.4 \mu s$. The noise parameters used here are taken from Ref.~\cite{de2010universal} and have been used for recent simulations such as in Ref.~ \cite{lei2017decoherence}.

\subsubsection{Static Magnetic fields}
\begin{figure}[!!htbp]
    \centering
    \begin{subfigure}{}
        \centering
        \includegraphics[width=3in]{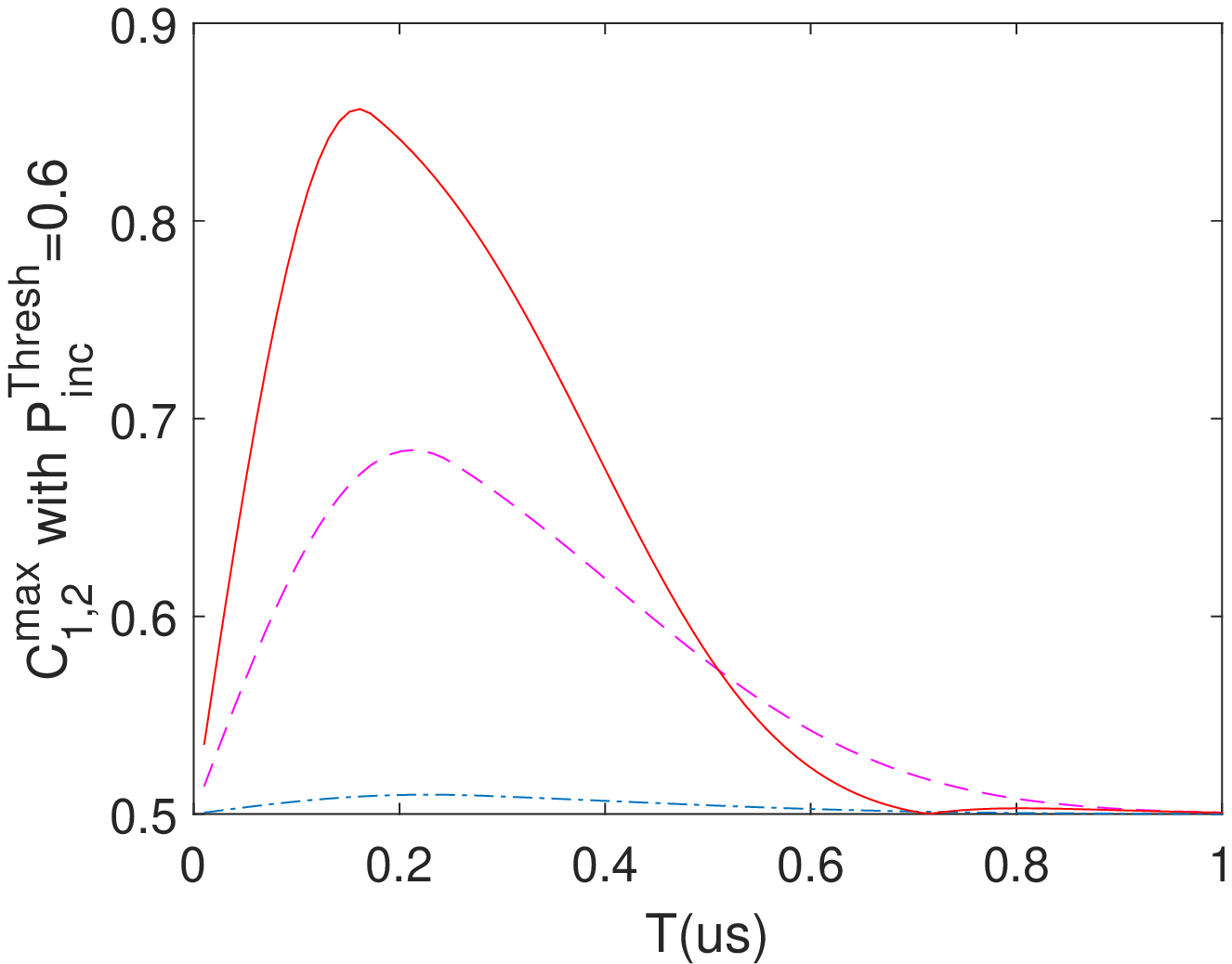}
    \end{subfigure}%
    \caption{(Color Online) Maximum Confidence $C_{0,1}^{\text{max}}$ as a function of total time $T$ (in $\mu$s) for constant magnetic fields of strengths 1$\mu$T (blue, dot-dashed line), 20 $\mu$T (magenta, dashed line) and 50$\mu$T (red, solid line) with $P_{\text{inc}}$ limited to $0.6$.}
    \label{static-thresha}
    ~
    \begin{subfigure}{}
        \centering
        \includegraphics[width=3in]{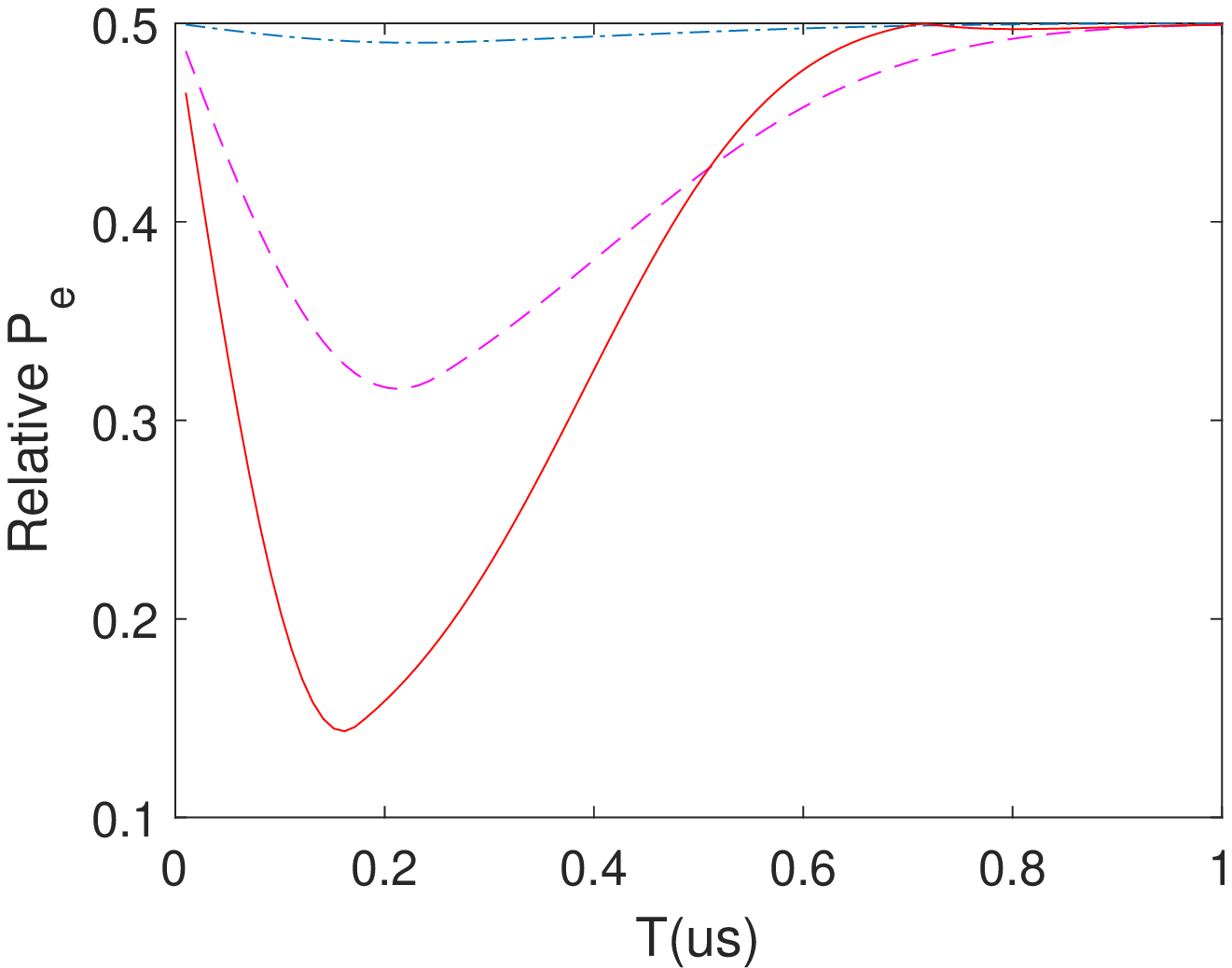}
        \caption{Color Online) Relative probability of error as a function of total time $T$ (in $\mu$s) for constant magnetic fields of strengths 1$\mu$T (blue, dot-dashed line), 20 $\mu$T (magenta, dashed line) and 50$\mu$T (red, solid line) with $P_{\text{inc}}$ limited to $0.6$.}
        \label{static-threshb}
    \end{subfigure}
\end{figure}

Let us first assume that we know the magnetic field to be detected perfectly, as in $B(t)=b$ and that there are no control fields applied to reduce the effect of dephasing of the single NV center. In this situation, we have $\nu=e^{-t/T_{2,\text{single}}^{*}}$ \cite{wang2012comparison}. This detection is typically performed in the pseudo-spin-1/2 single-quantum (SQ) subspace
of the NV- ground state, with the $|m_{s}=0\rangle$ and either the $|m_{s}=+1\rangle$ or $|m_{s}=-1\rangle$ spin state ($\Delta m_{s}=1$) employed. The accumulated phase factor in the presence of the magnetic field can be simply written as $\mu=e^{-i2\pi b\gamma T(\Delta m_{s})}$.
\begin{figure}[!htbp]
    \centering
    \begin{subfigure}{}
        \centering
        \includegraphics[width=3in]{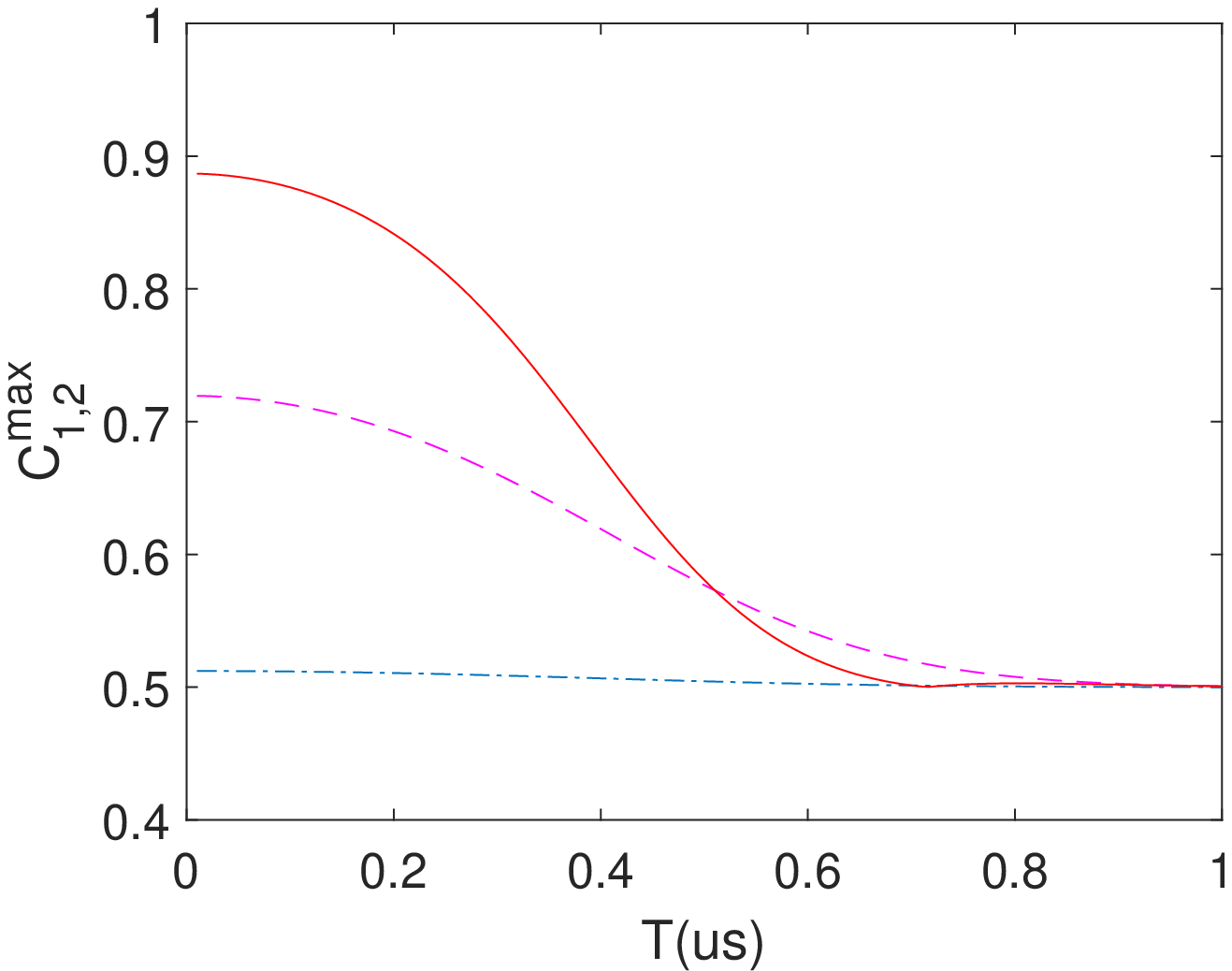}
    \end{subfigure}%
    ~
    \begin{subfigure}{}
        \centering
        \includegraphics[width=3in]{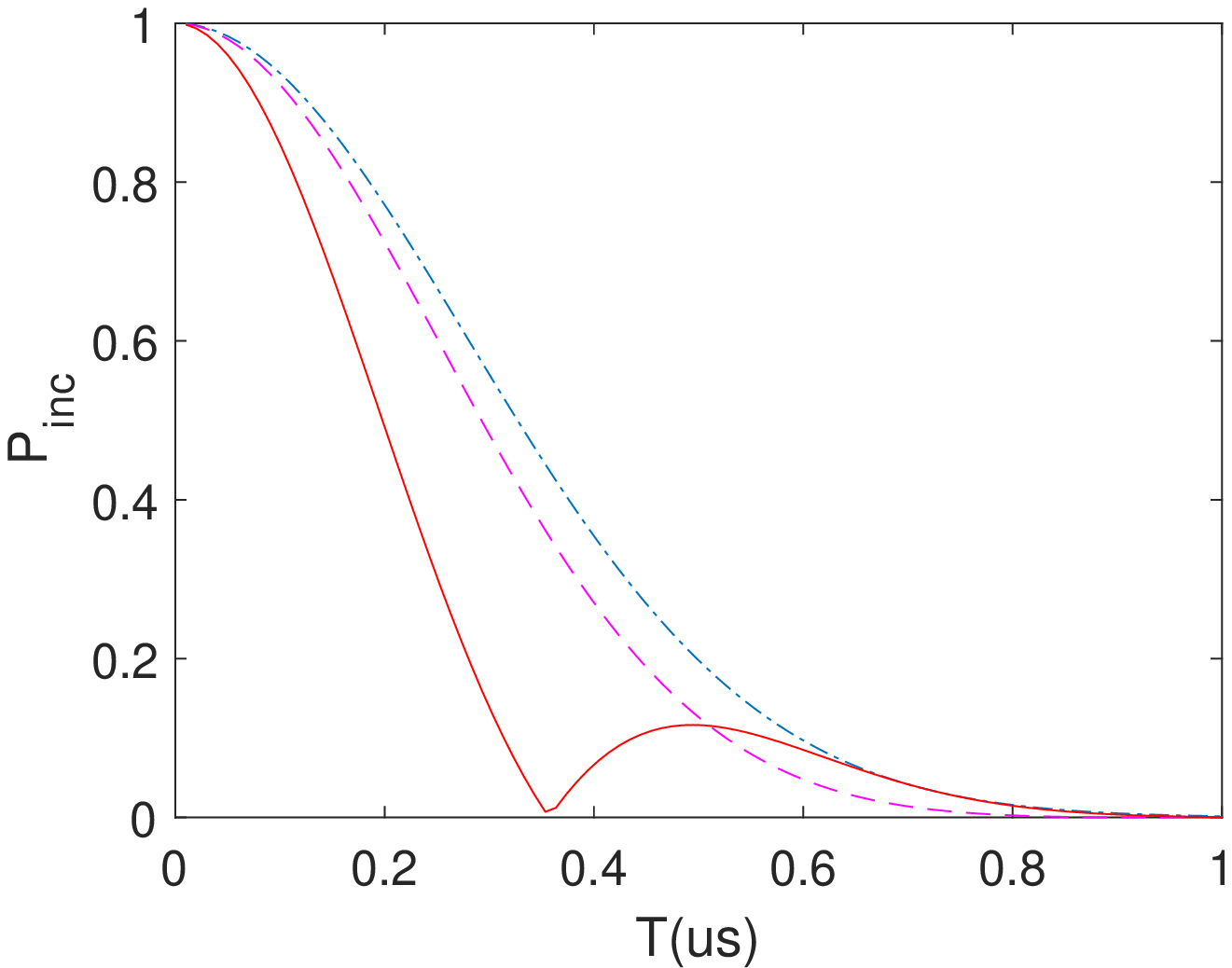}
        \caption{(Color Online) (a) Maximum Confidence $C_{1,2}^{\text{max}}$ as a function of total time $T$ (in $\mu$s) for constant magnetic fields of strengths 1$\mu$T (blue, dot-dashed line), 20 $\mu$T (magenta, dashed line) and 50$\mu$T (red, solid line). (b) Corresponding probability of inconclusive results $P_{\text{inc}}$.}
        \label{static-simple}
    \end{subfigure}
\end{figure}
Using these expressions alongside the discrimination formalism developed, we can find the confidence of our detection alongside the corresponding probability of inconclusive result as a function of the interaction time [see Fig.~\ref{static-simple}]. A simple glance at the graphs shows that our confidence is dependent on the magnetic field strength and it decays due to the effect of decoherence. Furthermore, the optimal value of $P_{\text{inc}}$ is extremely high in the beginning, which makes practical detection impossible, while at later times, it decays to zero meaning that the optimal measurement is again the von-Neumann strategy. A simple way to approach detection then could be to develop a weighted function between confidence and inconclusive rate and to use some median values as optimal detectors, or to limit the value of $P_{inc}$ to some standard value depending on the experimental resources available. The latter has been performed in Fig.~\ref{static-thresha}. Here we have set an upper bound threshold on $P_{\text{inc}}^{\text{thresh}}=0.6$. The benefit of our approach can also be seen from Fig.~\ref{static-threshb} where we have plotted the relative probability of error corresponding to Fig.~\ref{static-thresha}. A direct comparison with the measurements in \cite{chaudhry2015detecting} and we see an approximate $\sim$50$\%$ decrease in the detection error probability.

A more realistic scenario is that we do not know the value of the magnetic field precisely. Instead, the field $b$ follows some probability distribution which we assume to be $P(b)=\frac{1}{\sqrt{2\pi \sigma_{b}}}e^{-(b-b_{0})^{2}/2\sigma_{b}^{2}}$. In other words, we suspect that the magnetic field is around some average $b_{0}$ and we are trying to detect it. For this case
\begin{eqnarray}
\mu=\int P(b)e^{-i2\pi \gamma b T}=e^{-i2\pi \gamma b_{0} T} e^{-2\pi^{2}\gamma^{2}T^{2}\sigma_{b}^{2}},
\end{eqnarray}
while the decoherence factor $\nu$ remains the same. We can then find the corresponding measurements operators again that  maximise confidence as a function of $T$.

\begin{figure}[!htbp]
    \centering
    \begin{subfigure}{}
        \centering
        \includegraphics[width=3in]{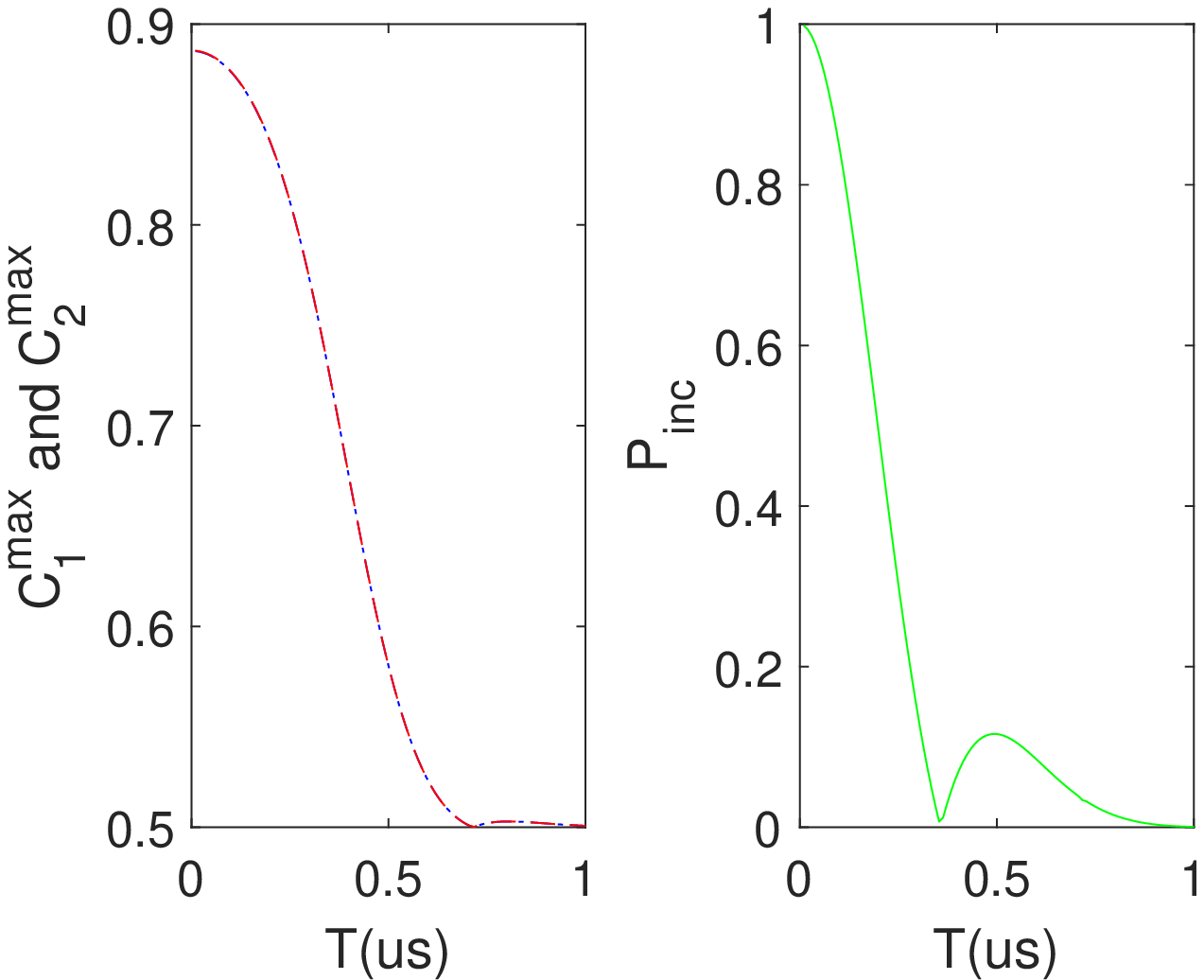}
        \caption{Maximum Confidence of detection of 50$\mu$T magnetic field with distribution $\sigma_{b}=1\mu\textrm{T}$ for $C_{1}^{\text{max}}$(red, dashed line) and $C_{2}^{\text{max}}$(blue, dot-dashed line); for this figure both are equal.}
        \label{gauss-const1}
    \end{subfigure}%
    ~
    \begin{subfigure}{}
        \centering
        \includegraphics[width=3in]{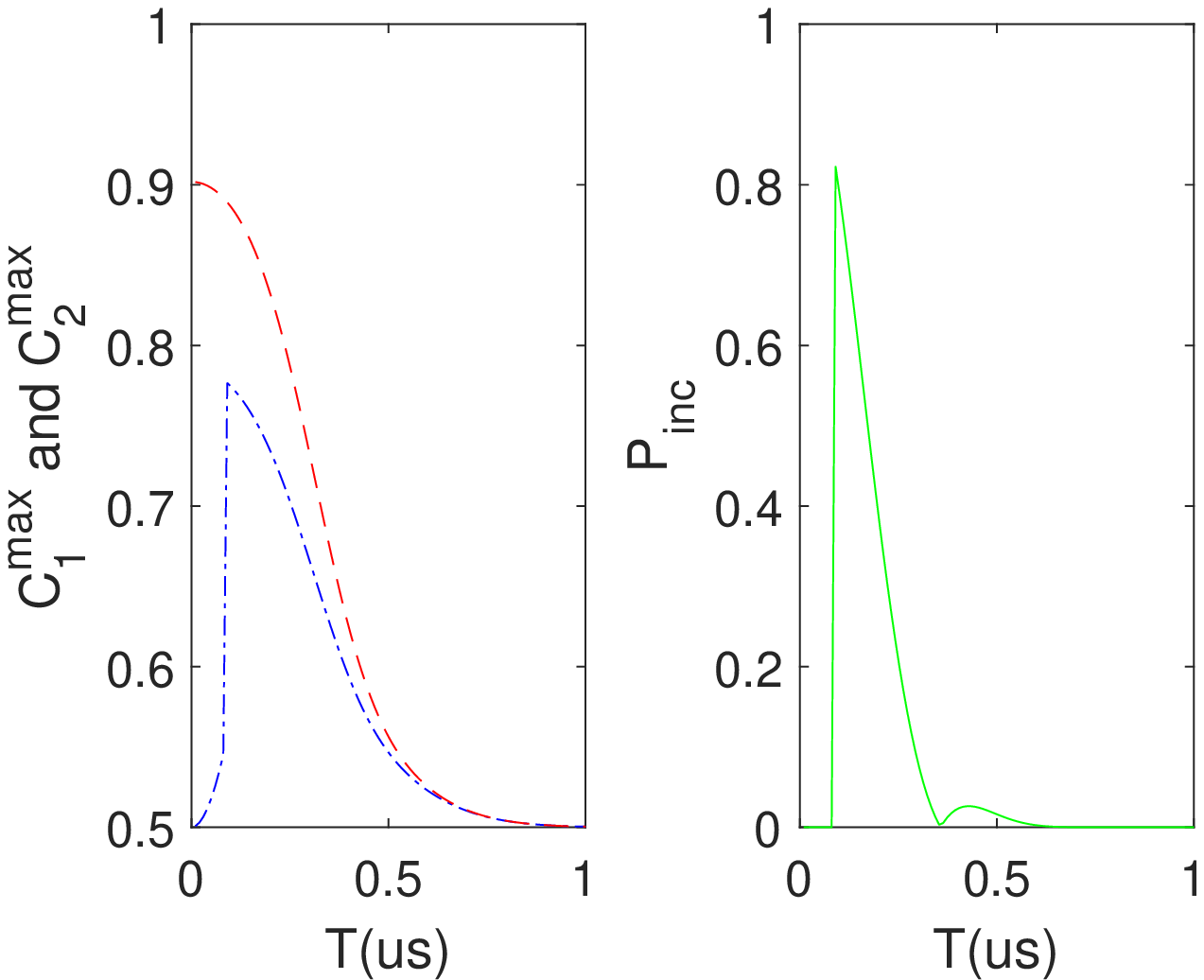}
        \caption{Confidence of detection of 50$\mu$T magnetic field with distribution $\sigma_{b}=25\mu\textrm{T}$ for $C_{1}^{\text{max}}$(red, dashed line) and $C_{2}^{\text{max}}$(blue, dot-dashed line)}
        \label{gauss-const2}
    \end{subfigure}%
    
    ~
    \begin{subfigure}{}
        \centering
        \includegraphics[width=3in]{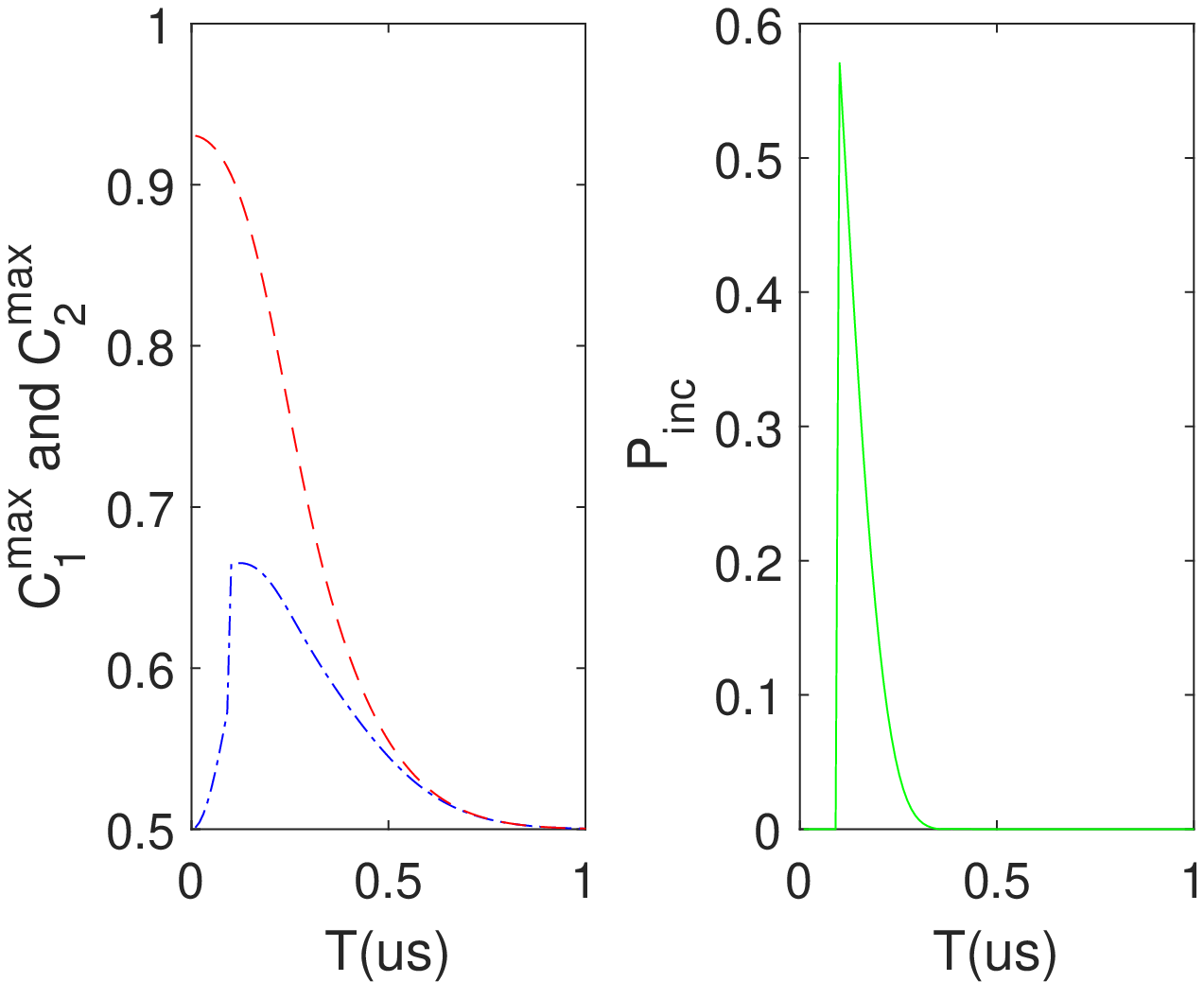}
        \caption{Confidence of detection of 50$\mu$T magnetic field with distribution $\sigma_{b}=50\mu\textrm{T}$ for $C_{1}^{\text{max}}$(red, dashed line) and $C_{2}^{\text{max}}$(blue, dot-dashed line)}
        \label{gauss-const3}
    \end{subfigure}
    
\end{figure}

From Figs.~\ref{gauss-const1}-\ref{gauss-const3}, we can see that the confidences $C_{0}^{\text{max}}\neq C_{1}^{\text{max}} $. This  illustrates the point that in general the maximum confidence measurement detection operators do not give exactly the same confidence. For the case of increasing $\sigma_{b}$, this disparity increases, meaning that the detectors become more biased towards detecting one outcome with more confidence than the other. 

\subsubsection{Oscillating Wave Magnetometry} 
\begin{figure}[!htbp]
    \centering
    \begin{subfigure}{}
        \centering
        \includegraphics[width=3in]{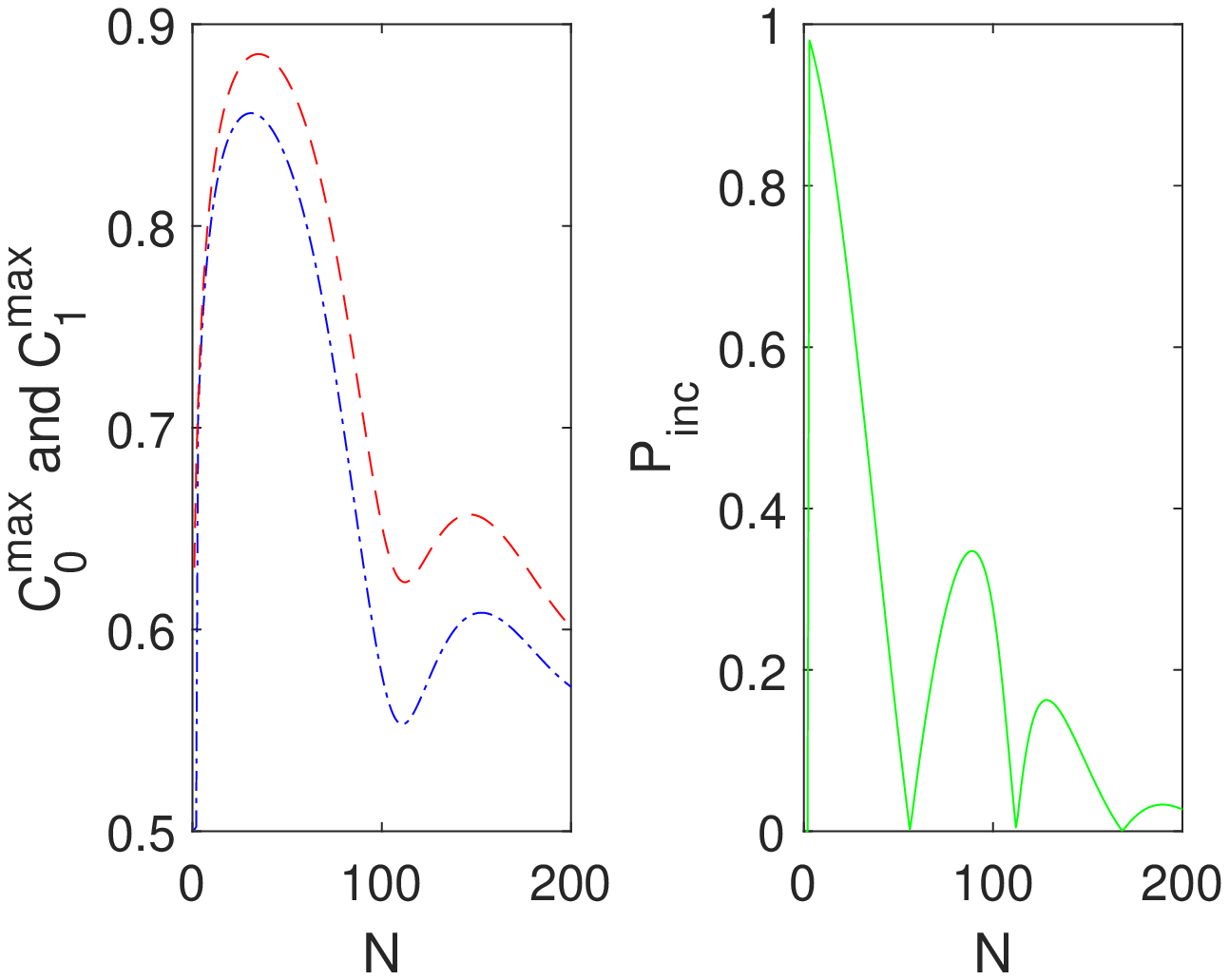}
        \caption{Confidence of detection of an oscillating magnetic field $B(t)=b_{0} \cos(2\pi f t)$ with $b_{0}=1\mu\textrm{T}$,$\sigma_{b}=0.2\mu\textrm{T}$ and $f=1$ MHz for $C_{0}^{\text{max}}$(red, dashed line) and $C_{1}^{\text{max}}$(blue, dot-dashed line)}
        \label{CPMG-cos1}
    \end{subfigure}%
    ~
    \begin{subfigure}{}
        \centering
        \includegraphics[width=3in]{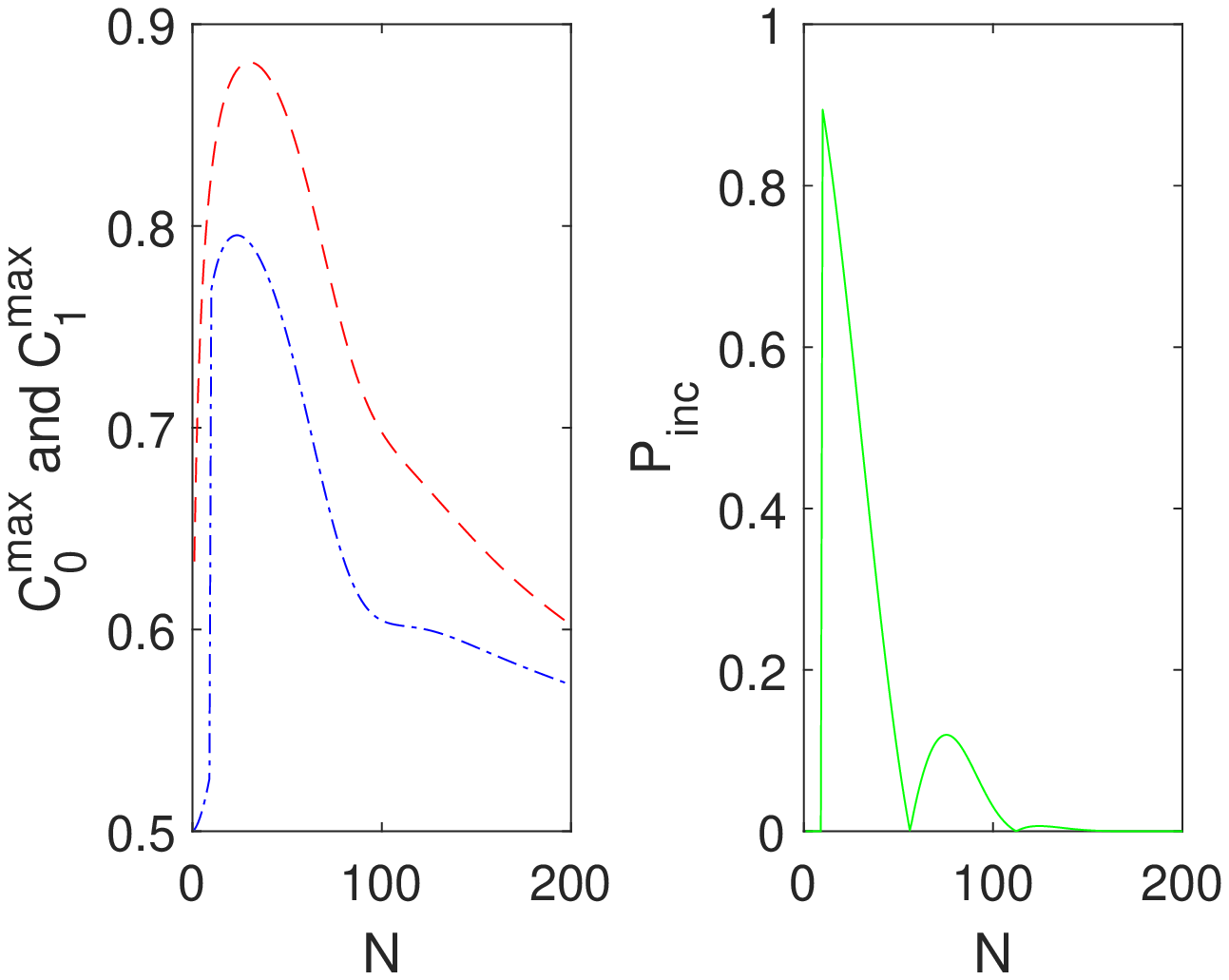}
        \caption{Same as Fig.~\ref{CPMG-cos1}, except that now $\sigma_b = 0.4$ $\mu$T.}
        \label{CPMG-cos2}
    \end{subfigure}
    ~
    \begin{subfigure}{}
        \centering
        \includegraphics[width=3in]{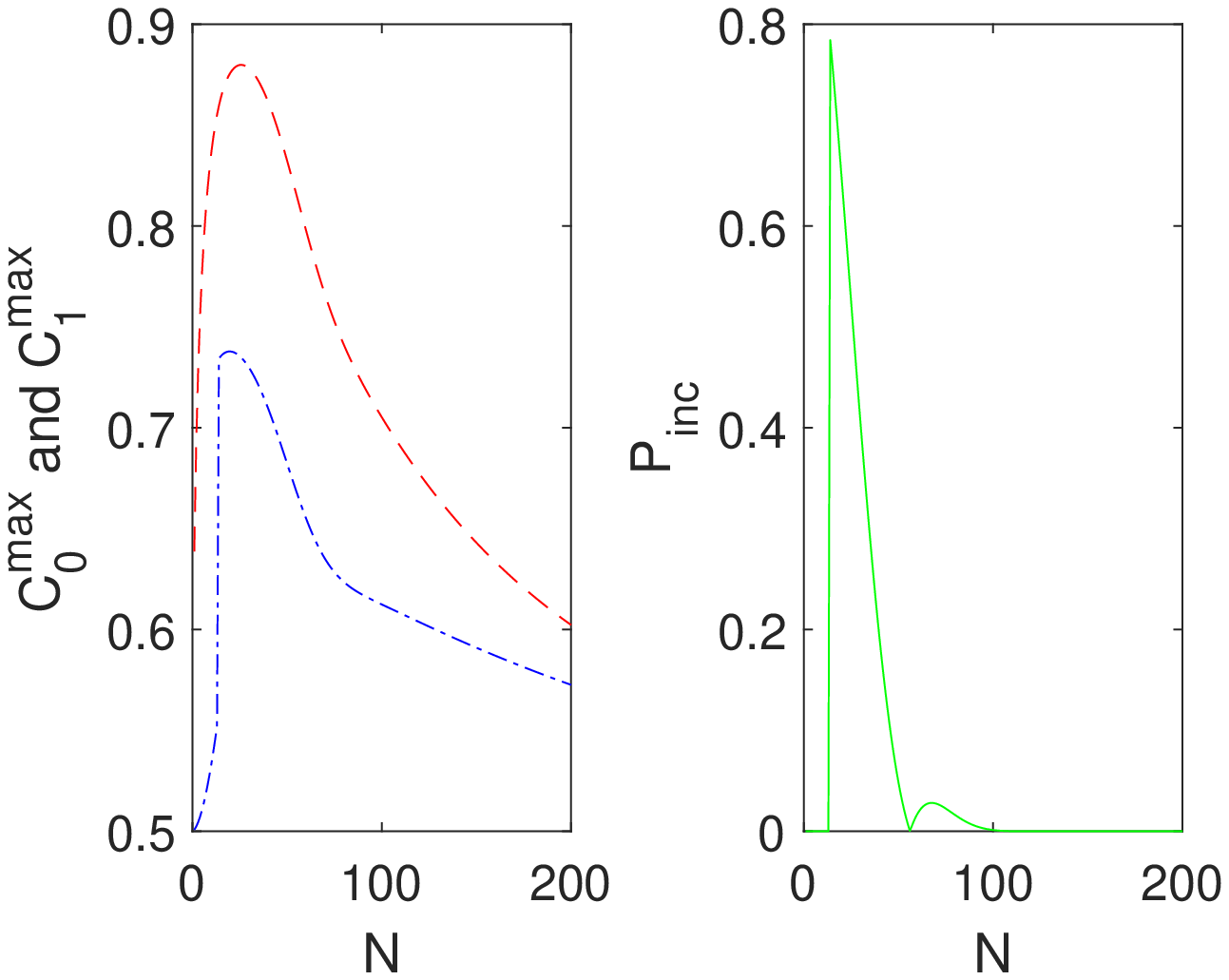}   
        \caption{Same as Fig.~\ref{CPMG-cos1}, except that now $\sigma_b = 0.6$ $\mu$T.}
        \label{CPMG-cos3}
    \end{subfigure}
    
\end{figure}

In the simulations performed thus far, the limitation has been decoherence and the strength of the magnetic field to be detected. One of the methods to reduce decoherence is dynamic decoupling \cite{viola1998dynamical,viola1999dynamical,de2010universal,naydenov2011dynamical,ryan2010robust,du2009preserving,biercuk2009optimized} which involves the use of pulse sequences such as the Carr-Purcell-Meiboom-Gill (CPMG) pulse sequence. The applied pulses keep on flipping the sign of the NV center-spin bath interaction Hamiltonian, which, if done in rapid succession, can effectively eliminate the effect of the spin bath. Unfortunately, the effect of a static field is also removed. However, for an oscillating field the direction of the magnetic field also continues to change direction allowing its detection using dynamic decoupling protocols, provided that we apply the pulses at (or near) the nodes of the magnetic field to allow the accumulation of the phase difference. In particular, the CPMG sequence is characterised by $[U(\tau/2)R(\pi)U(\tau)R(\pi)U(\tau/2)]^{N/2}$, meaning that we allow the NV center to evolve freely for time $\tau/2$, then apply a control pulse defined by $R(\pi)=e^{-i\pi\sigma_x /2}$, followed by further evolution for time $\tau$. This cycle is repeated $N/2$ times, where $N$ corresponds to the number of pulses applied. The magnetic field to be detected has the form $B(t)=b_0 \cos{(2\pi ft)}$, and we assume a Gaussian distribution for the amplitude. To ensure that the pulses are applied at the nodes we set $\tau=1/2f$. We then find $\mu=e^{-i2N\gamma b_0/f}e^{-2N^2\gamma^2\sigma_{b}^2/f^2}$. The computation for $\nu$ is more challenging. In the presence of pulses, $\nu=\langle \exp[-i\int_{0}^{T}\xi(t)B_d(t)dt]\rangle$ where $\xi(t)$ can assume values of $+1$ and $-1$ - the switching of the sign account for the action of a pulse. It can then be shown that $\nu=e^{-\kappa^2W(T)}$ with $W(T)=\int_{0}^{T}e^{-Rs}p(s)ds$, where $R=1/\tau_c$ and $p(s)=\int_{0}^{T-s}\xi(t)\xi(t+s) dt$. With full derivations available in Refs.~\cite{wang2012comparison,chaudhry2014utilizing,chaudhry2015detecting}, the expression of $\nu$ can be calculated allowing us to find the number of pulses $N$ that would maximize the confidence of the measurement. The results for the case where $f=1$ MHz is given in Figs.\ref{CPMG-cos1}-\ref{CPMG-cos3}. As we can see from these figures, the effect of the CPMG pulses reduces the effect of decoherence, making the values of $\nu$ remain much closer to one. This consequently leads to considerably high levels of confidence in our measurements for $b_0=1\mu T$ compared to the static case. However, similar to the static case, a greater spread in the magnetic field corresponds to our detectors becoming biased towards giving one result with better confidence.

\subsection{Ensembles}
Having investigated the detection using a single NV center in detail, let us now try to determine how our results change when we use an ensemble of NV centers with an outstretched parameter $p=1$ to make the same detection measurements. We now have $T^{*}_{2,\text{ensemble}}=0.4\mu \text{s}$ \cite{bauch2018ultralong}.
\subsubsection{Static fields}
\begin{figure}[!htbp]
    \centering
    \begin{subfigure}{}
        \centering
        \includegraphics[width=3in]{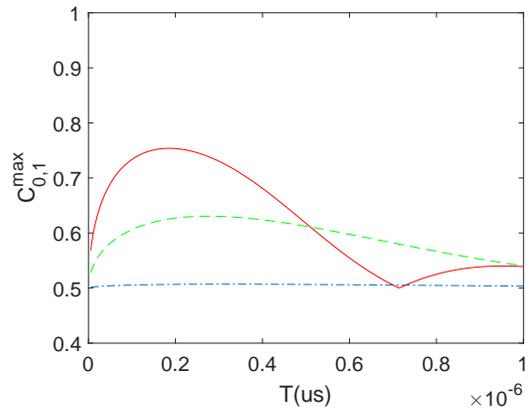}
    \end{subfigure}%
    ~
    \begin{subfigure}{}
        \centering
        \includegraphics[width=3in]{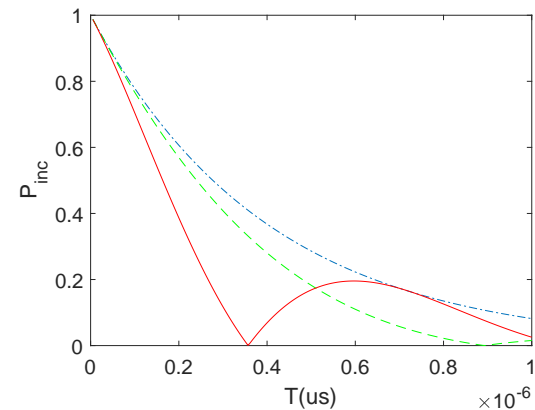}
        \caption{(Color Online) (a) Maximum Confidence $C_{0,1}^{\text{max}}$ as a function of total time $T$ (in $\mu$s) for constant magnetic fields of strengths 1$\mu$T (blue, dot-dashed line), 20 $\mu$T (green, dashed line) and 50$\mu$T (red, solid line).(b) Corresponding probability of inconclusive results $P_{\text{inc}}$.}
        \label{ens-static-simple}
    \end{subfigure}
\end{figure}

Just as before for the static case, we use $B(t)=b$, and define $\nu=e^{-t/T_{2,\text{ensemble}}^{*}}$ with the same $\mu$. Results are shown in Fig.~\ref{ens-static-simple}.
We can see that, unlike the static measurement for the case of a single NV center, the maximum confidence reached is lower for the ensemble case. However, since the exponent in the exponential decoherence factor is linear for the ensemble case (and not quadratic like the single case), the confidences achieved are maintained for a longer period of time.

\begin{figure}[!htbp]
    \centering
    \begin{subfigure}{}
        \centering
        \includegraphics[width=3in]{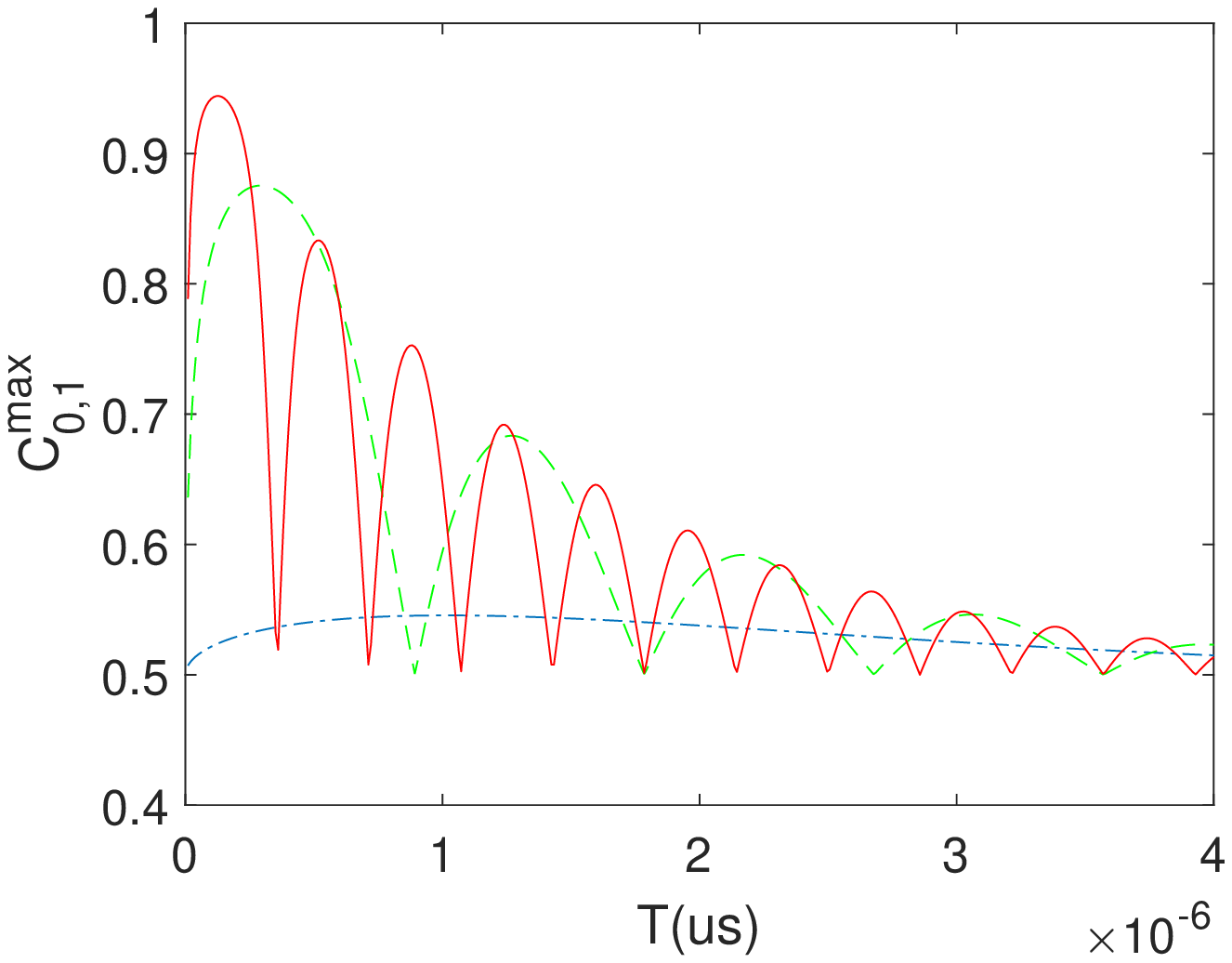}
    \end{subfigure}%
    ~
    \begin{subfigure}{}
        \centering
        \includegraphics[width=3in]{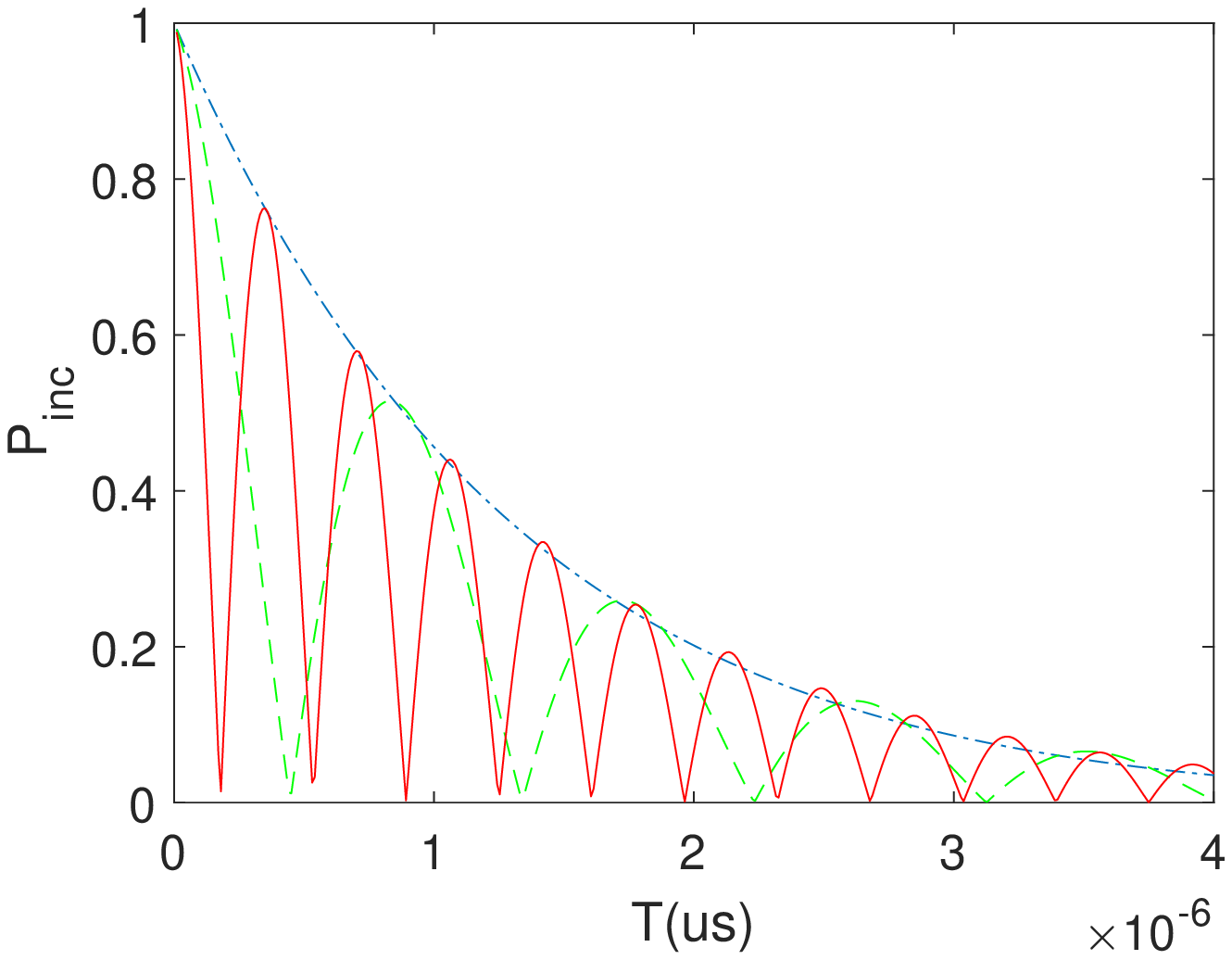}
        \caption{(Color Online) (a) Maximum confidence $C_{0,1}^{\text{max}}$ as a function of total time $T$ (in $\mu$s) for constant magnetic fields of strengths 1$\mu$T (blue, dot-dashed line), 20 $\mu$T (green, dashed line) and 50$\mu$T (red, solid line). (b) Corresponding probability of inconclusive results $P_{\text{inc}}$.}
        \label{ens-static-DQ}
    \end{subfigure}
\end{figure}

A way to improve our ability to detect static magnetic fields is to use double quantum magnetometery (DQ) which works by first preparing an equal superposition of the $m_{s}=|+1\rangle$ and $|m_{s}=-1\rangle$ states (for example, $|+_{DQ}\rangle=(1/\sqrt{2})(|+1\rangle+|-1\rangle)$), which is then exposed to a magnetic field during the free precession interval followed by the final population being mapped to the $|m_{s}=0\rangle$ state. Use of DQ magnetometery allows for several advantages. In the previously used single quantum (SQ) basis, nonmagnetic noise sources such as temperature fluctuations, electric-field noise, and inhomogeneous strain may also contribute to spin dephasing. However, values of $T_{2,ens}^{*}$ in the DQ basis are insensitive to these common-mode noise sources. Furthermore, the phase accumulation doubles for DQ magnetometry since $\Delta m_{s}=+2$ although the dephasing time also doubles resulting in $T_{2,DQ}^{*}\approx T_{2,SQ}^{*}/2$ (for our simulations, we have used $T^{*}_{2,DQ}=1.3\mu s$). However, for isotopically purified, low nitrogen diamond, $T_{2,DQ}^{*}$ can actually exceed the coherence time in the SQ basis \cite{bauch2018ultralong}, thereby allowing us to benefit from the doubled phase factor and increased coherence time. This is demonstrated in Fig.\ref{ens-static-DQ}, where we can see a much higher level of confidence is achieved as compared to Fig.~\ref{ens-static-simple}.

We may ask why we did not examine the DQ magnetometery for single NV center. The reason is simply that as proven in Ref.~ \cite{de2009electron}, $T_{2,single}^{*}$ in the SQ basis is simply not affected by the zero-frequency signals such as spatially inhomogeneous
magnetic fields, electric fields, strain, or g-factors, which is why changing the basis of measurement simply results in doubling the dephasing time and the phase accumulation, which does not result in a significant change in detection confidence. 

\begin{figure}[!htbp]
    \centering
    \begin{subfigure}{}
        \centering
        \includegraphics[width=3in]{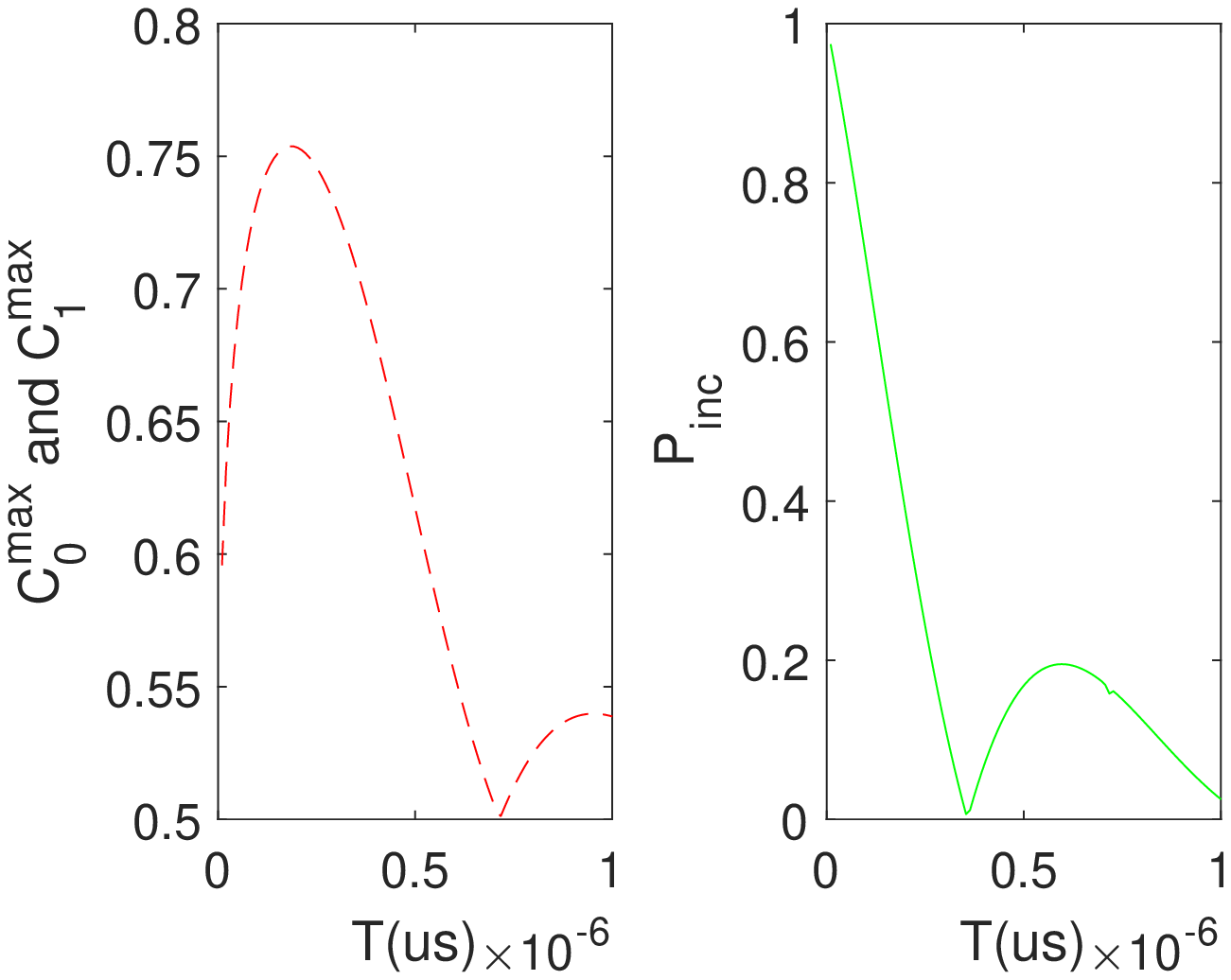}
        \caption{Confidence of detection of 1$\mu$T magnetic field with distribution $\sigma_{b}=1\mu\textrm{T}$ for $C_{0}^{\text{max}}$ (red, dashed line) and $C_{1}^{\text{max}}$(blue, dot-dashed line); for this figure both are equal.}
        \label{gauss-ens1}
    \end{subfigure}%
    ~
    \begin{subfigure}{}
        \centering
        \includegraphics[width=3in]{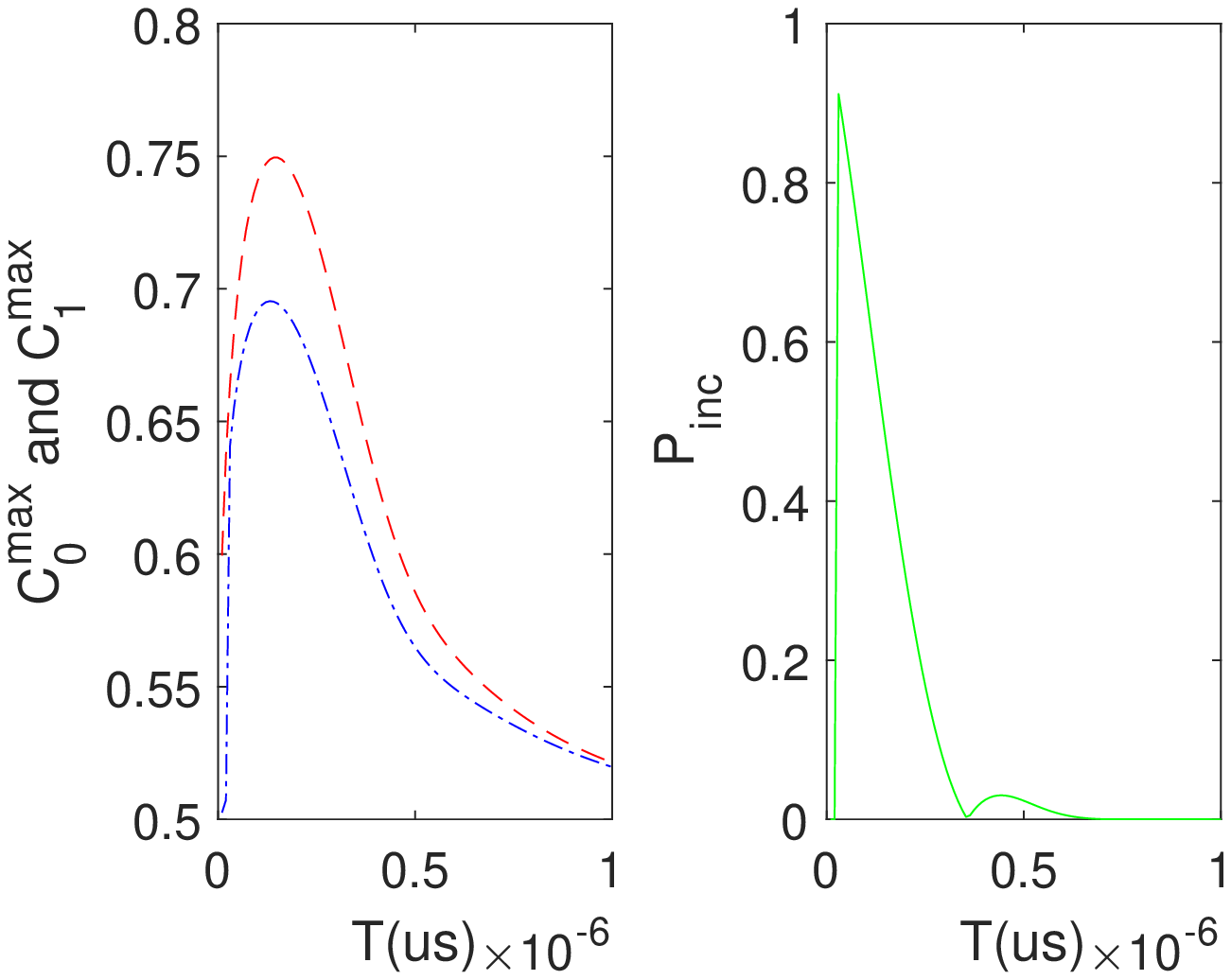}
        \caption{Confidence of detection of 25$\mu$T magnetic field with distribution $\sigma_{b}=25\mu\textrm{T}$ for $C_{0}^{\text{max}}$ (red, dashed line) and $C_{1}^{\text{max}}$ (blue, dot-dashed line).}
        \label{gauss-ens2}
    \end{subfigure}%
    ~
    \begin{subfigure}{}
        \centering
        \includegraphics[width=3in]{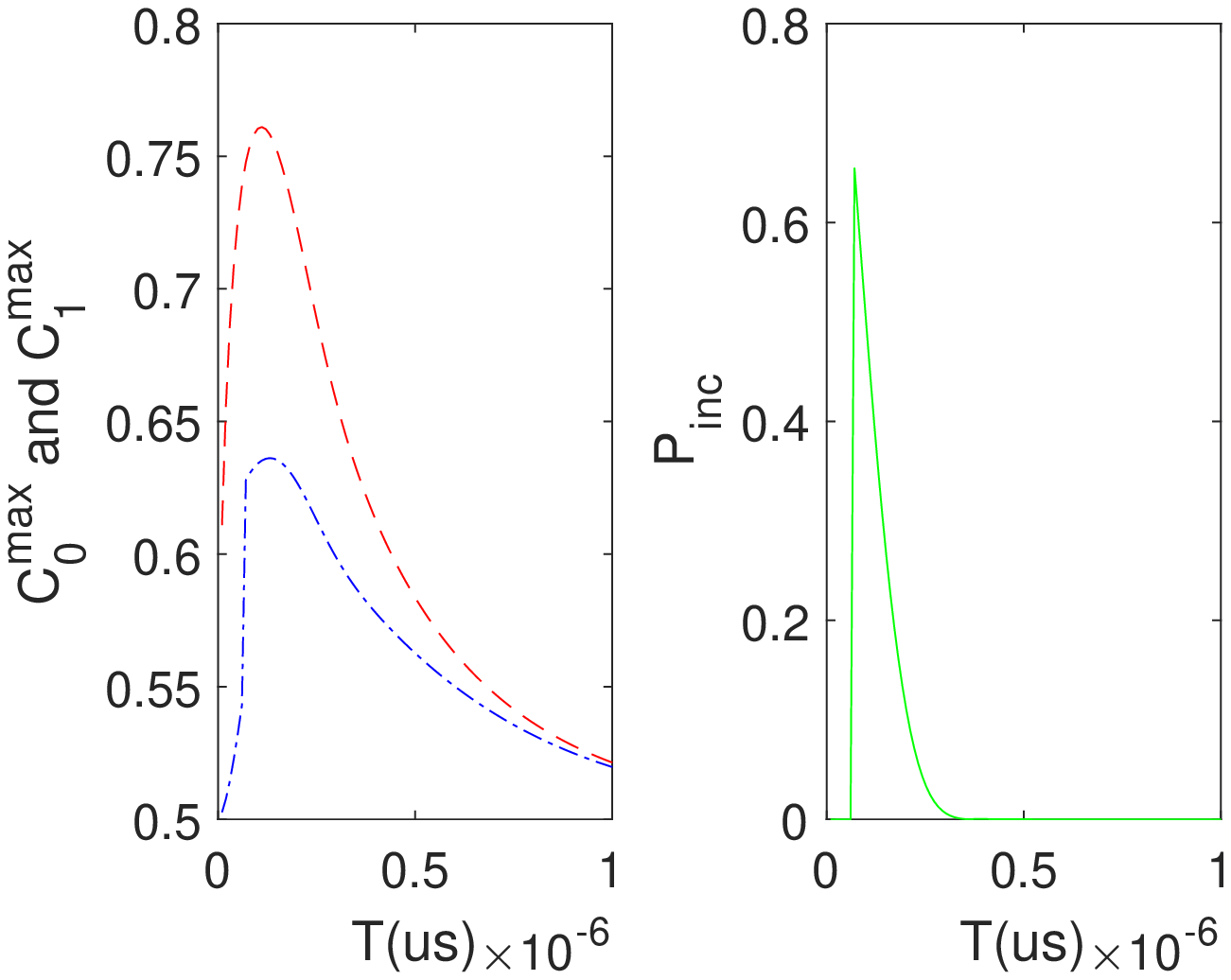}
        \caption{Confidence of detection of 50$\mu$T magnetic field with distribution $\sigma_{b}=50\mu\textrm{T}$ for $C_{0}^{\text{max}}$ (red, dashed line) and $C_{1}^{\text{max}}$ (blue, dot-dashed line).}
        \label{gauss-ens3}
    \end{subfigure}
    
\end{figure}

Our analysis for Gaussian static fields follows as outlined before, however we note that the simulations are performed here in the SQ basis simply for better comparisons with the results from the single NV centers. The results are plotted in Figs.~ \ref{gauss-ens1}-\ref{gauss-ens3}, where the similar trend for lower overall confidence for longer periods of time is observed.

\subsubsection{Oscillating fields}
\begin{figure}[!htbp]
    \centering
    \begin{subfigure}{}
        \centering
        \includegraphics[width=3in]{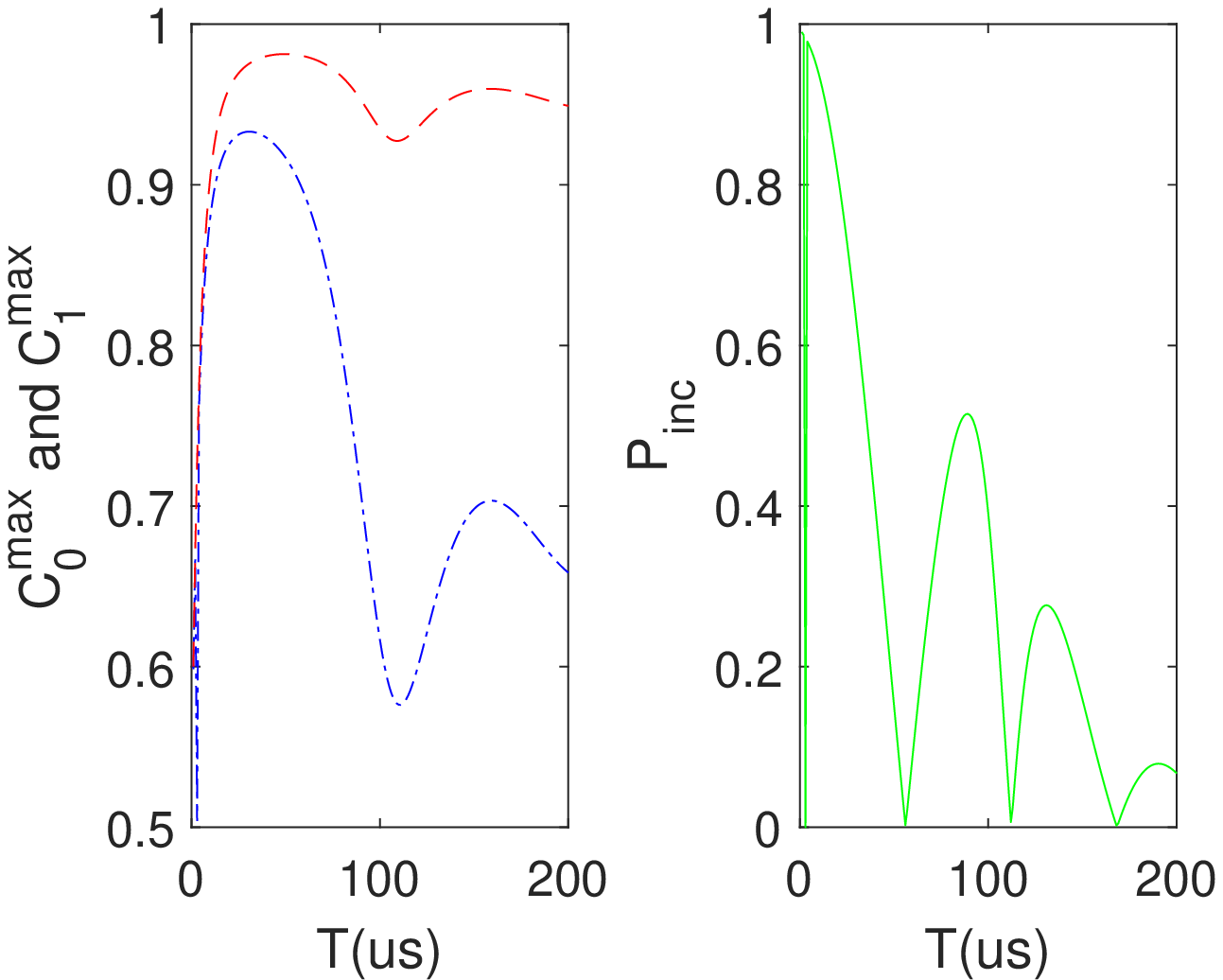}
        \caption{Confidence of detection of Oscillating magnetic field $B$($t$)=$b_{0}\cos(2\pi f t)$ with $b_{0}=1\mu\textrm{T}$, $\sigma_{b}=0.2\mu\textrm{T}$ and $f = 1$ MHz for $C_{0}^{\text{max}}$ (red, dashed line) and $C_{1}^{\text{max}}$ (blue, dot-dashed line)}
        \label{CPMG-cos-ens1}
    \end{subfigure}%
    ~
    \begin{subfigure}{}
        \centering
        \includegraphics[width=3in]{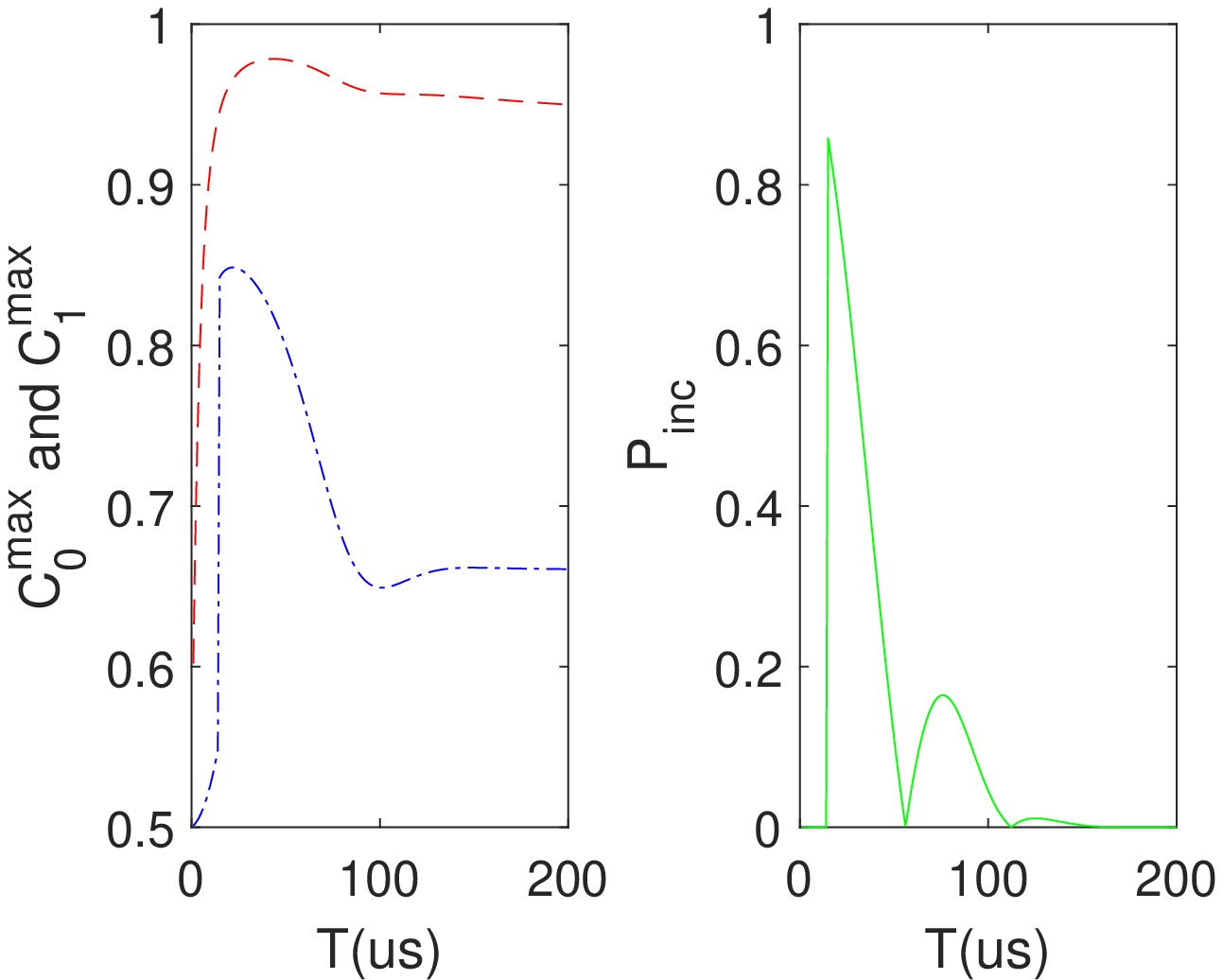}
        \caption{Same as Fig.~\ref{CPMG-cos-ens1}, except that now $\sigma_b = 0.2$}
        \label{CPMG-cos-ens2}
    \end{subfigure}
    ~
    \begin{subfigure}{}
        \centering
        \includegraphics[width=3in]{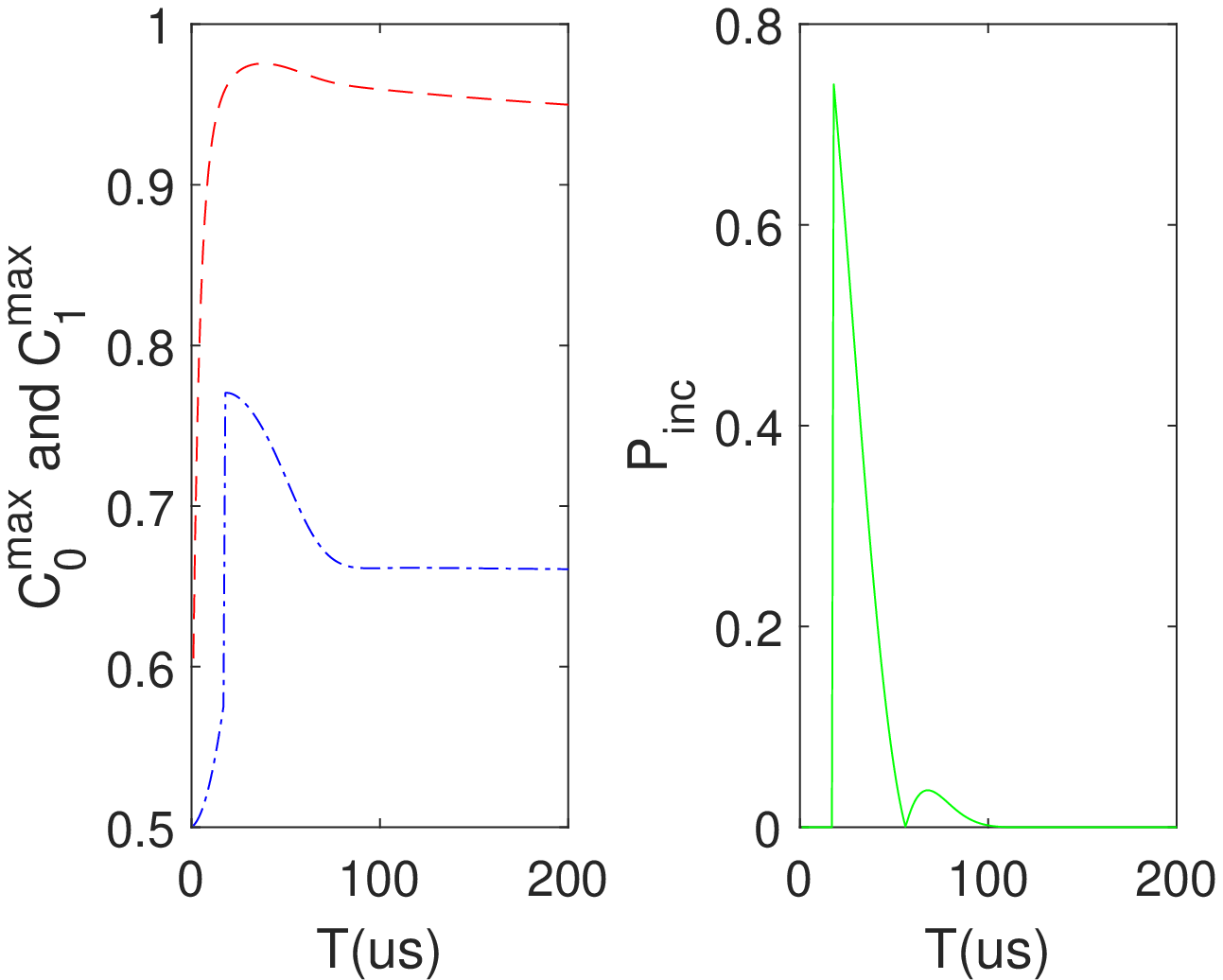}   
        \caption{Same as Fig.~\ref{CPMG-cos-ens1}, except that now $\sigma_b = 0.6$.}
        \label{CPMG-cos-ens3}
    \end{subfigure}
 
\end{figure}

We now apply the CPMG sequence to detect a sinusoidal magnetic field. The calculation of the decoherence under the pulse sequence is done in a slightly different manner. To do so, we use the general formalism outlined in \cite{barry2020sensitivity,pham2012enhanced} giving us $\nu=exp[-(\frac{N^{1-s}}{2T_2f})^p]$, where $s$ is set by the noise spectrum of the decohering bath. For our use case of an electronic spin bath, this can be approximated as $2/3$ \cite{de2009electron}. $T_2$ is the coherence time equal to 53 $\mu$s, and, as before, $f$ represents the frequency of the magnetic field and $N$ is the number of pulses applied. Our results are shown in Figs.\ref{CPMG-cos-ens1}-\ref{CPMG-cos-ens3} where we see an actual increase in the confidence as well as longer lasting confidence measurements as a function of time.

\section{Implementation of generalized measurements}
We have provided until now a general framework which benefits from the ability of generalized measurements to give inconclusive results in order to improve the confidence with which we can measure the system. While these measurements provide great flexibility in characterizing the probabilities of the different measurement outcomes, we need to understand how to actually perform these generalized measurement. Our generalized measurement operators are three rank-1 matrices($\Pi_0,\Pi_1$ and $\Pi_?$) in a 2-dimensional Hilbert space. The key towards implementation is then to use Neumark's theorem \cite{herzog2010generalized} to realize our generalized measurement operaters as projective measurements in an extended 3-dimensional Hilbert space.

We begin by introducing a normalized ancilla bit $\ket{2}$ with $\langle 0 | 2 \rangle = \langle 1 | 2 \rangle = 0$. With the constraints on the POVM operators, 
\begin{eqnarray*}
\Pi_0+\Pi_1+\Pi_?=&&\ketbra{\pi_0}{\pi_0}+\ketbra{\pi_1}{\pi_1}+\ketbra{\pi_3}{\pi_3}\\\nonumber
=&&\ketbra{0}{0}+\ketbra{1}{1}=I_2,
\end{eqnarray*} 
where we have used the normalized states $\ket{0}$ and $\ket{1}$, and noted that each generalized measurement operator is rank $1$ so that we can write them as $\Pi_0=\ketbra{\pi_0}{\pi_0}$, $\Pi_1=\ketbra{\pi_1}{\pi_1}$,  and $\Pi_?=\ketbra{\pi_3}{\pi_3}$. It is then possible to determine projectors $P_k=\ketbra{e_k}{e_k}$ of the form
\begin{eqnarray}
\ket{e_k}=\ket{\pi_k}+c_k\ket{2}\;\;\; \textrm{with}\;\;\; \braket{e_k}{e_{k'}}=\delta_{kk'}.
\end{eqnarray}
We can further choose the phase of the ancillary $\ket{2}$, without loss of generality, such that $c_0$ is real and positive. This results in
\begin{eqnarray}
c_0=(1-\braket{\pi_0}{\pi_0})^{1/2},\;\;\;\; c_{1(2)}=-\frac{\braket{\pi_0}{\pi_{1(2)}}}{c_0}
\label{thec}
\end{eqnarray}
It should be apparent that now the measurements using the operators Tr($\rho_j \Pi_k$) for $j=0,1$ are equivalent to performing $\bra{\pi_k}\rho_j\ket{\pi_k}=\bra{e_k}\rho_j\ket{e_k}$, meaning that the generalized measurements correspond to projections along $\ket{e_k}$. In order to actualize these projections in the extended Hilbert space,  we can use a unitary transformation on a set of orthonormal vectors $\ket{0},\ket{1} \textrm{ and } \ket{2}$ which are related by
\begin{eqnarray}
\textrm{Tr}(\rho_j\Pi_2)=\bra{0}U\rho_j 
U^{\dagger{}}\ket{0}\;\;\;\nonumber
\textrm{for}\;\;U^{\dagger{}}\ket{0}=\ket{e_2},\\\nonumber
\textrm{Tr}(\rho_j\Pi_1)=\bra{1}U\rho_j 
U^{\dagger{}}\ket{1}\;\;\;
\textrm{for}\;\;U^{\dagger{}}\ket{1}=\ket{e_1},\\ \nonumber
\textrm{Tr}(\rho_j\Pi_?)=\bra{2}U\rho_j 
U^{\dagger{}}\ket{2}\;\;\;
\textrm{for}\;\;U^{\dagger{}}\ket{2}=\ket{e_0},\\\nonumber
\end{eqnarray}
where we have set up the transformation with the intention that the inconclusive result will be measured by projection onto the auxillary state. Finally, the resulting unitary transform using Eq.~\eqref{thec} is simply
\begin{eqnarray}
U=
\left( \begin{array}{ccc}
\braket{\pi_2}{0} & \braket{\pi_2}{1} & c_2 \\
\braket{\pi_1}{0} & \braket{\pi_1}{1} & c_1 \\
\braket{\pi_0}{0} & \braket{\pi_0}{0} & c_0 \\
\end{array} \right).
\end{eqnarray}

Since the NV center is a 3 level quantum system with only two levels typically used in magnetometry, the extension of the Hilbert space is straightforward. Empty levels can be used to extend the Hilbert space \cite{franke2001generalized,herzog2010generalized}. This, coupled with the fact that any three-dimensional unitary operator
can be decomposed into a product of at most
three two-dimensional unitary operators acting in
two-dimensional subspaces of the total Hilbert space, means that two-dimensional unitary operators can be used to implement our detection operators.

\section{Conclusion}
In conclusion we have used NV centers to detect both constant and oscillating magnetic fields. Our aim has been to maximize the confidence of the detection. We have used ensembles and single NV centers to qualitatively see how differently they perform as detectors. We have used techniques like DQ magnetometry and dynamic decoupling to improve our results while providing a model for how these generalized measurements could be implemented experimentally. This work should be useful for the detection of weak magnetic fields using NV centers.


\providecommand{\noopsort}[1]{}\providecommand{\singleletter}[1]{#1}%

\end{document}